\documentclass[12pt]{iopart}

\usepackage{graphicx}
\usepackage{iopams}
\usepackage[breaklinks=true, colorlinks=true, linkcolor=blue, urlcolor=blue, citecolor=blue]{hyperref}
\usepackage{dsfont} 

\usepackage{verbatim}

\begin{document}

\title[Not-Post-Peierls  compatibility under noisy channels]{Not-Post-Peierls compatibility under noisy channels}

\author{Andr\'es F Ducuara$^{1,2}$, Cristian E Susa$^{1,2,3}$ and John H Reina$^{1,2}$} 
\address{$^{1}$Departamento de F\'isica, Universidad del Valle, 760032 Cali, Colombia}
\address{$^{2}$Centre for Bioinformatics and Photonics---CIBioFi, Calle 13 No. 100-00, Edificio 320 No. 1069, 760032 Cali, Colombia}
\address{$^{3}$Departamento de F\'isica y Electr\'onica, Universidad de C\'ordoba, Carrera 6 No. 76-103, Monter\'ia, Colombia} 
\ead{andres.ducuara@correounivalle.edu.co
\\ cristiansusa@correo.unicordoba.edu.co
\\ john.reina@correounivalle.edu.co }
\vspace{10pt}
\begin{indented}
\item[]February 2017
\end{indented}

\begin{abstract}
The Pusey-Barrett-Rudolph (PBR) theorem deals with the realism of the quantum states. It establishes that every pure quantum state is real, in the context of quantum ontological models. Specifically, by guaranteeing the property of not-Post-Peierls ($\neg$PP) compatibility (or antidistinguishability) for a particular set of states $P$, together with the ad hoc postulate known as Preparation Independence Postulate (PIP), the theorem establishes that these two properties imply the $\psi$-onticity (realism) of the set of all pure states. This PBR result has triggered two particular lines of research: On the one hand, it has been possible to derive similar results without the use of the PIP, although at the expense of implying weaker properties than $\psi$-onticity. On the other hand, it has also been proven that the property of $\neg$PP compatibility alone is an explicit witness of usefulness for the task known as conclusive exclusion of states. In this work, we explore the $\neg$PP compatibility of the set of states $P$, when $P$ is under the interaction of some noisy channels, which would consequently let us identify some noisy scenarios where it is still possible to perform the task of conclusive exclusion of states. Specifically, we consider the set $P$ of $n$-qubit states in interaction with an environment by means of i) individual and  ii) collective couplings. In both cases, we analytically show that the phenomenon of achieving $\neg$PP compatibility, although reduced, it is still present. Searching for an optimisation of this phenomenon, we report numerical experiments up to $n=4$ qubits. For individual qubit-noise coupling, be it a bit, a phase, or a bit--phase flip noise, the numerical search exhibits the same response to all the different noisy channels, without improving the analytical expectation. In contrast, the collective qubit-noise coupling leads to a more efficient numerical display of the phenomenon, and a variety of noise channel-dependent behaviours emerge.  Furthermore, by adopting a slight modification of the definition of $\psi$-onticity, from being a property exclusively of sets of pure states to be a property of general sets of states (now $\rho$-onticity), in particular of the considered noisy set $P$, our results can also be seen as addressing the $\rho$-ontic realism of the set of $n$-qubit states $P$ under the considered noisy channels.
\end{abstract}

%
\vspace{2pc}
\submitto{\JPA}
%
%
%

\section{Introduction}

\label{intro}
Quantum mechanics is a proven successful theory, however, there still exist conceptual ambiguities about  its interpretation of \textit{reality}, i.e., a reality that is independent of the measurements undertaken in the laboratory. For instance, some of these striking quantum features are exhibited by the Einstein-Podolsky-Rosen (EPR) argument \cite{EPR1935}, which deals with the property of \emph{completeness} (and a generalisation of it called \emph{$\psi$-onticity} \cite{HS2010}), the Bell's theorem, which deals with the property of \emph{locality} \cite{Bell1964}, and the Kochen-Specker theorem, which deals with the property of \emph{non-contextuality} \cite{KS1967}. These properties can be incorporated and dealt within the same grounds by means of the \emph{quantum ontological models} (QOMs) \cite{Bell1964, HS2010}. Within this formalism, it is clear that quantum theory is a nonlocal-contextual theory \cite{JL2016}, however, it is still not known whether it can be referred to as $\psi$-ontic (i. e. the probability functions over the so-called hidden variables associated to different quantum states have non-overlapping supports), or $\psi$-epistemic (not $\psi$-ontic). Even though $\psi$-epistemic models are able to reproduce several aspects of quantum mechanics \cite{TM2007}, the Pusey-Barrett-Rudolph (PBR) ``no-go''  theorem \cite{PBR2012} aims to discard them. Specifically, by considering the property of not-Post-Peierls ($\neg$PP) compatibility of a particular set of states $P$, together with the so-called {\it preparation independence postulate} (PIP) (which means that the attached probability function that relates a pure-separable state with these hidden variables is factorisable), the PBR theorem establishes that these two properties imply that the set of all pure states must be modelled by a $\psi$-ontic QOM (discarding $\psi$-epistemic models). In short, that all pure states are real ($\psi$-ontic). This PBR result has triggered two particular lines of research. On the one hand, there have been derived other no-go theorems aiming to discard $\psi$-epistemic models as the Hardy \cite{Hardy2013}, and Colbeck-Renner \cite{CR2013} theorems. However, as the PBR result, all of these theorems have to make extra additional assumptions in order to derive their results \cite{ML2014} (and it has been proven that if we go around those assumptions, it is still possible to build $\psi$-epistemic models \cite{CE1, CE2, Shane2016}). Ideally, we should look forward to discard $\psi$-epistemic models without invoking any additional assumption. In this regard, a particular class of models that can be completely discarded are the so-called  \emph{maximally $\psi$-epistemic} models \cite{MPE1, MPE2, MPE3, MPE4}, which has already been experimentally verified \cite{EXP12013, EXP22015, EXP32016}. On the other hand, it has also been proven that the property of $\neg$PP compatibility (alone) turns out to be useful for the task known as the \emph{conclusive exclusion of states} \cite{CES2014, CES2015, CES2016}. 

In this work, we explore the $\neg$PP compatibility of the set $P$, when $P$ is under the influence of some noisy channels, which consequently lets us address the efficiency of the task of conclusive exclusion of states after the noises. The way in which a system interacts with its environment signals its quantum dynamics \cite{Z2003, Mazzola2010, RSF2014, Modi2012, MSDBR2017, SR2010, SR2012, RQJ2002, Kroja2006}. Many interesting features and applications of quantum systems come from their interactions with the environment, such as quantum decoherence and collective effects \cite{Z2003, RSF2014, SR2010, SR2012, RQJ2002, Kroja2006}, and different entanglement and quantum correlations phenomena \cite{Mazzola2010, Modi2012, MSDBR2017}, crucial aspects at the foundations and applications of quantum mechanics. A simple form of studying the effects of the environment over the system is through the noise channels formalism \cite{NC2010, OQS2002}, which  includes bit flip, phase flip, and bit-phase flip channels \cite{NC2010}. All of these can elegantly be described within  the operator-sum (Kraus) representation \cite{NC2010, Kraus1983}. We explicitly investigate the effects of bit, phase, and bit-phase flip channels on the $\neg$PP compatibility of the set of $n$-qubit states $P$, and their implications on the task of conclusive exclusion of states. Furthermore, by adopting a slight modification of the definition of $\psi$-onticity, from being a property exclusively of sets of pure states, to be a property of general sets of states (so now $\rho$-onticity), the noisy $P$ in particular, these results can also be seen as addressing the $\rho$-ontic realism of the set of $n$-qubit states $P$ after the interaction with the considered noisy channels.

This work is organised as follows: We start by making a concise review of the QOM formalism and the PBR argument (the interested reader is encouraged to check the comprehensive reviews of this and similar results in \cite{JL2016, ML2014}). We then comment on two particular lines of research that the PBR result has triggered. Next, we move into the main part of this work, where we address the $\neg$PP compatibility of the set $P$ when it is under the action of some noisy channels, particularly, we consider the following two situations for the system-environment interaction. Firstly, we assume that only one qubit encounters the action of the noisy channels whilst the remainder qubits are unaffected, in an \emph{individual} way. Secondly, we explore the case of more than one qubit being changed by the noises, in a \emph{collective} way. In both cases, we analytically show that the phenomenon of achieving $\neg$PP compatibility, although reduced, is still present. We support our analytical findings by carrying out a numerical approach which we run up to four qubits. Finally, we present a discussion on the implications of our results on the task of conclusive exclusion of states and on the generalised $\rho$-ontic realism of the set of $n$-qubit states $P$.

\section{Quantum Ontological Models}

We start by introducing the concept of Prepare-Measure (PM) fragments of quantum theory \cite{ML2014}, to then consider ontological models associated with such quantum fragments, or Quantum Ontological Models (QOMs) \cite{HS2010, ML2014}.

\subsection{Prepare-Measure (PM) fragments of quantum theory}

A Prepare-Measure (PM) fragment of quantum theory is a structure $(\mathds{H}, P, M)$ \cite{ML2014} where: $\mathds{H}$ is a finite-dimensional Hilbert space, $P$ is a set of an arbitrary amount of quantum states, say $P=\{ \rho\} \subseteq D(\mathds{H})$, with $D(\mathds{H})$ the set of density matrices on $\mathds{H}$, and $M$ is a set of an arbitrary amount of measurements $m$ (POVMs), being $m\in M \subseteq M^A$ a single POVM, and $M^A$ the set of all POVMs. The POVM elements $E_k$'s, $m =\{  E_k\}\in M$, satisfy the conditions $\sum_k E_k=\mathds{1} $ and $E_k \succcurlyeq 0$.  When $P=D(\mathds{H})$ and $M=M^A$, the fragment $(\mathds{H}, P, M)$ ends up covering the whole amount of states and measurements for a given $\mathds{H}$. The quantity of interest is the probability of obtaining outcome $k$ when measuring $m$ and the system is prepared in the quantum state $\rho$ which is given by the Born rule as ${\rm Tr} (E_k\rho)$ \cite{NC2010}. Let us now introduce the concept of ontological model for general PM fragments of quantum theory. 

\subsection{Ontological models for PM fragments of quantum theory}

\begin{figure}[h!]
\centerline{\includegraphics[scale=0.8]{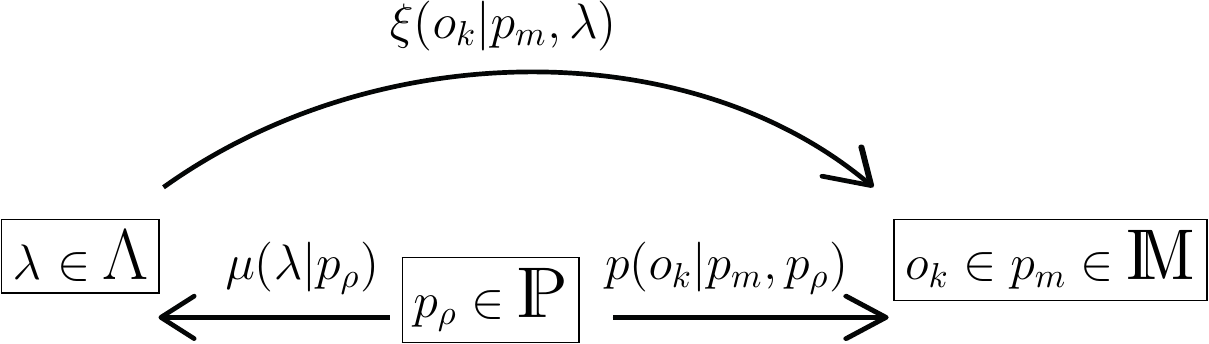}}
\caption{General scheme of an ontological model for the PM fragment $(\mathds{H}, P, M)$. The conditional probabilities $\mu$, $\xi$, and $p$ relate the different sets of the model $(\Lambda, \mathds{P}, \mathds{M})$.} 
\label{fig:fig1}
\end{figure}

We now consider an ontological model for the PM fragment $(\mathds{H}, P, M)$, or Quantum Ontological Model (QOM), as the structure composed by the sets $(\Lambda, \mathds{P}, \mathds{M})$ whose elements are denoted as $\lambda \in \Lambda$, $p_{\rho} \in \mathds{P}$, $p_m \in \mathds{M}$, and conditional probability functions $\{\mu(\lambda|p_{\rho})\}, \{\xi(o_k|p_m, \lambda)\}, \{p(o_k|p_m, p_{\rho})\})$ that relate these sets as depicted in \autoref{fig:fig1}.

\noindent
We now further detail the structure of a QOM \cite{HS2010, ML2014}. $\mathds{P}$ is the set of \emph{state preparation procedures}, $p_{\rho} \in \mathds{P}$ is a procedure to experimentally generate the quantum state $\rho \in P$. In general, different state preparation procedures might end up generating the same quantum state, so there is actually a set $\mathds{P}_{\rho}\subsetneq \mathds{P}$ of different preparation procedures generating $\rho$. $\mathds{M}$ is the set of \emph{measurement preparation procedures}, $p_m \in \mathds{M}$ is a procedure to experimentally implement the measurement $m=\{E_k\}\in M$.  This preparation procedure $p_m$ is in turn composed by the outcomes of the experiment $o_k$. In general, different measurement preparation procedures might end up generating the same measurement, so there is actually a set $\mathds{M}_m\subsetneq \mathds{M}$ of different measurement procedures generating $m$. $\Lambda$ is the so-called set of ontic states, with $\lambda \in \Lambda$ an ontic state.

\noindent
We also have conditional probability functions that relate the elements of these sets as it is depicted in \autoref{fig:fig1}, where $\mu(\lambda|p_{\rho})\equiv \mu_{\rho}(\lambda)$, with $\mu_{\rho}: \Lambda \rightarrow [0, 1]$ a probability function, and similarly for the other functions. The function $\mu(\lambda|p_{\rho})$ represents the probability for the system to be in the ontic state $\lambda$, when the system  has been prepared in $\rho$ with procedure $p_{\rho}$ (different procedures $p'_{\rho}$  generating the same state $\rho$, might have a different function $\mu(\lambda|p'_{\rho})$). The function $\xi(o_k|p_m, \lambda)$ represents the probability of obtaining outcome $o_k$ when measuring $m$ with procedure $p_m$ and the ontic state of the system being $\lambda$ (different procedures $p'_m$ generating the same state $m$, might have a different function $\xi(o_k|p'_m, \lambda)$). The function  $p(o_k|p_m, p_{\rho})$ represents the probability of obtaining outcome $k$ from measurement $m$ prepared with $p_m$, when preparing the state $\rho$ by means of the procedure $p_{\rho}$. Furthermore, we have that these conditional probability functions satisfy the so-called total probability law which reads:
\begin{eqnarray}
p(o_k|p_m, p_{\rho})=\int_{\Lambda} d\lambda \mu(\lambda|p_{\rho}) \xi(o_k|p_m, \lambda).
\label{eq:TPL}
\end{eqnarray}
We are interested into reproducing quantum mechanical probabilities (given by the Born rule) with these OMs, we therefore impose the identity:
\begin{eqnarray}
p(o_k|p_m, p_{\rho})=\int_{\Lambda} d\lambda \mu(\lambda|p_{\rho}) \xi(o_k|p_m, \lambda):={\rm Tr} (E_k\rho),
\label{eq:BR}
\end{eqnarray}
where $p_{\rho}$ and $p_m$ are particular state and measurement preparation procedures. In general, this could also be achieved by means of other procedures, say $p(o_k|p'_m, p'_{\rho})$ with $o_k \in p'_m$, however, for the sake of notation we will only assume particular procedures $p_{\rho}, p_m$ throughout the rest of the document, with the caveat that it is actually sets of different preparation procedures $\mathds{P}_{\rho}$ and $\mathds{M}_{m}$. Let us now move into a particular class of these QOMs.

\subsection{$\psi$-ontic QOMs}

\begin{figure}[h!]
\centerline{\includegraphics[scale=0.7]{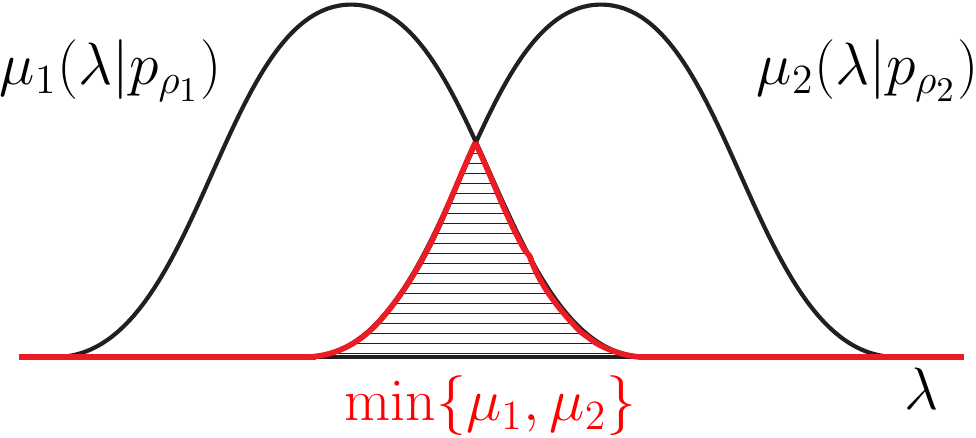}}  
\vspace{-0.2cm}
\caption{Schematic representation of the overlap between probabilities $ \mu_{1}$ and $ \mu_{2}$.} 
\label{fig:fig2}
\end{figure}

Given $P=\{ \rho_{\vec x} \}$, with respective functions $\{ \mu_{\vec x}(\lambda)\}$ (where $\vec x$ stands as a counter which is going to be useful later), we define the {\it overlap between probabilities} as
\begin{eqnarray}
 w(\{ \mu_{\vec x}\}):=\int_{\Lambda}   {\rm min}_{\vec x} \{   \mu_{\vec x}( \lambda) \}d \lambda.
 \label{eq:overlap}
\end{eqnarray}
We have a $\psi$-ontic QOM  if and only if  $w(\{ \mu_{\vec x}\})=0$ (See \autoref{fig:fig2}) \cite{ML2014}. In other words, this definition encapsulates the idea of not overlapping functions $\{ \mu_{\vec x}\}$. We also say that $\{ \rho_{\vec x} \}$ are ontologically distinct.

\noindent
These $\psi$-ontic QOMs are important because they generalise the so-called \emph{$\psi$-complete} QOMs,  which are complete in the EPR's sense \cite{HS2010}  \footnote{The QOM formalism here introduced can actually be more formally addressed by considering a measure-theoretic approach \cite{ML2014}, which turns out mandatory when considering \emph{$\psi$-complete} models. However, the treatment here presented turns out enough for the purposes of this work, because we will be focus on upper bounding the magnitude $w$ with a function $\sigma$ which invokes quantum mechanics alone, and this result, in addition of being derived from the measure-theoretic approach, can also be written from the present approach as we are going to further detail in Lemma 2 and Appendix B.}. This is the  origin of the conceptual wording `$\psi$-ontic realism' or `$\psi$-onticity'. It should also be noted that this definition depends on the considered set $\{ \rho_{\vec x} \}=P\subseteq D(\mathds{H})$, where $P$ is in principle, an arbitrary set of density matrices. In this regard, we are going to address this definition as a property of sets of pure states \cite{ML2014} (as it is usually addressed), however, we are also going to consider it as a property of general sets of states (sets with mixed states included) which we will address as $\rho$-onticity whenever is the case. A QOM that is not $\psi$-ontic  is called $\psi$-epistemic. The PBR argument deals with $\psi$-ontic QOMs and with a postulate termed ``preparation independence''.

\subsection{Preparation Independence Postulate}

Given the set of two quantum states $P''=\{ \rho_{x_i} \} \subseteq D(\mathds{H})$, with functions $\mu_{x_i}(\lambda_{x_i})$, $x_i\in\{0,1\} $, let us consider the set of states:
\begin{eqnarray}
P=\{\rho_{\vec x}\}  \subset D(\mathds{H} ^{\otimes n}), \hspace{0.3cm} \rho_{\vec x}=\bigotimes^n_{j=1} \rho_{x_j},
\label{eq:BS}
\end{eqnarray}
with $ \vec {x} :=(x_1,...,x_n)$, $x_j\in \{0, 1\}$. Then, for the new functions $\mu_{\vec x}(\vec{\lambda})$, with $ \vec  \lambda = ( \lambda_1,..., \lambda_{n})\in \Lambda^n $, we can assume the factorisation:
\begin{eqnarray}
 \mu_{\vec x}(\vec \lambda)= \mu_{x_1}(\lambda_1)...\mu_{x_n}( \lambda_{n}),
\label{eq:PIP} 
\end{eqnarray}
which is the so-called {\it preparation independence postulate} (PIP). In Appendix A, we present a derivation of this postulate for the particular case of local measurements. We next establish a relation between the set of states (\autoref{eq:BS}) and the original set of states $P''=\{\rho_0, \rho_1\}$. 

\subsubsection{Lemma 1:}

Given $P''=\{\rho_0, \rho_1\} \subset D(\mathds{H})$, assuming PIP, and considering the states given by \autoref{eq:BS}, the overlap between the probabilities (\autoref{eq:overlap}) becomes:
\begin{eqnarray}
\nonumber w(\{ \mu _{\vec x} \})= w(\{\mu_0, \mu_1\} )^{n}.
\end{eqnarray}
In other words, if there exists an $n$ such that the set of states $P$ (\autoref{eq:BS}) is $\psi$-ontic ($w(\{ \mu _{\vec x} \})=0$), then the original set $P''$ of two states is $\psi$-ontic as well ($w(\{\mu_0, \mu_1\} )=0$). A derivation of this result is given in Appendix B.

\subsection{Not-Post-Peierls Compatibility}

This is a property  originally used in another context instead of the QOMs \cite{CFS2002}, and establishes the following: $P$ is \emph{Post-Peierls compatible} if:
\begin{eqnarray}
\nonumber \forall m, \exists  E _k\in m, \;\;\;\; \mathrm{s. t.:}\;\;\;\; \forall  \rho  \in P , \hspace{0.3cm} {\rm Tr}(E _k \rho) >0,
\end{eqnarray}
then the \emph{not-Post-Peierls ($\neg$PP) compatibility} (or PP incompatibility) property of $P$ becomes:
\vspace{-0.1cm}
\begin{eqnarray}
\exists m, \forall E _k\in m , \;\;\;\; \mathrm{s. t.:}\;\;\;\; \exists \rho_k \in P, \hspace{0.3cm}{\rm Tr}(E _k \rho_{k }) =0.
\label{eq:NPP}
\end{eqnarray}
To check that this property holds for a general $P$ system is a non trivial task \cite{CFS2002}. This property has also been called antidistinguishability \cite{ML2014}, and has also been proven to be a witness of usefulness for the task of conclusive exclusion of states \cite{CES2014}. Let us now connect  this property to  the $\psi$-ontic QOMs.

\subsubsection{Lemma 2:}

Let $P \subseteq D(\mathds{H})$, if $P$ satisfies $\neg$PP compatibility (for certain $m \in M$) (\autoref{eq:NPP}), then the QOM for the $PM$ fragment is $\psi$-ontic:
\begin{eqnarray}
\nonumber  \neg \mathrm{PP} \longrightarrow  \psi \mathrm{-ontic}.
\end{eqnarray}
This, in principle, allows us to determine the $\psi$-ontic realism in sets of states $P$ by means of the $\neg$PP compatibility. The proof of this lemma is given in Appendix C.  Next, we formulate the PBR argument.

\section{PBR Argument}

In this section we state the PBR theorem \cite{PBR2012}, followed by a numerical approach which originally aimed to outperform the analytical result, but rather ended up confirming its efficiency \cite{PBR2012}, however, we should bear in mind that this might had not been the case. This numerical approach will let us numerically explore the $\neg$PP compatibility for general sets of states. 

\subsection{PBR Theorem}

Given $P\subset D(\mathds{H})$ the set of all pure states, and accepting PIP, then $P$ satisfies $\psi$-onticity.

{\bf{Proof:}} Given the set of two arbitrary pure quantum states $P''=\{\rho_0, \rho_1\}  \subset D(\mathds{H})$, 
$\rho_{x_i}= \left|\psi(x_i)\right>\left<\psi(x_i)\right|$, it is always possible to rewrite them as
\begin{eqnarray}
\left|\psi(x_i)\right>=\cos\theta /2 \left| 0 \right>+(-1)^{x_i}\sin\theta /2 \left| 1\right>,
\label{eq:OS} 
\end{eqnarray}
where  $x_i \in \{0,1\}$,  $0\leq \theta <\frac{\pi}{2}$. 
In order to prove that they are ontologically distinct (\autoref{eq:overlap}) through $\neg$PP compatibility (\autoref{eq:NPP}), we can consider a new set of states as in \autoref{eq:BS}:
\begin{eqnarray}
\hspace{-1cm}P=\{ \rho_{\vec x} \}, \hspace{0.5cm} \rho_{\vec x}= \left| \Psi(\vec x)\right> \left <\Psi(\vec x)\right|, \hspace{0.5cm}  \left|\Psi(\vec x)\right>=\bigotimes_{i=1}^{n} \left|\psi(x_i)\right>,
\label{eq:pbrbigger}
\end{eqnarray}
with $\vec x =(x_1, x_2,...,x_n)$. Next, we need  to show that this new set $P$ satisfies $\neg$PP compatibility. For the set of states in \autoref{eq:pbrbigger}, PBR \cite{PBR2012} found an analytical bound which we address in what follows.

\subsubsection{Lemma 3}

Let $P=\{ \rho_{\vec x} \}$ be given by \autoref{eq:pbrbigger}; given $\theta$ there exists a $n$ such that $P$ is $\neg$PP compatible. That $n$ must satisfy \cite{PBR2012}:
\begin{eqnarray}
\theta \geq  2\arctan\left(2^\frac{1}{n}-1\right).
\label{eq:bound}
\end{eqnarray}
Using Lemma 2, the set of states $P$ defined by  \autoref{eq:pbrbigger} is $\psi$-ontic, and then by Lemma 1, the original pair of states (\autoref{eq:OS}) are ontologically distinct, and since they are two arbitrary pure states in $D(\mathds{H})$, we have that the whole set of pure states is $\psi$-ontic $\square$. 

In Appendix A, we show that the PIP is a natural assumption for pure-separable states and local measurements; however, since we cannot guarantee it for general states and measurements (in particular, for the measurements used by PBR \cite{PBR2012} in the proof of $\neg PP$ compatibility in Lemma 3), the PIP  remains with the status of an ad hoc postulate. The PBR argument \cite{PBR2012}, applies to all pure states, regardless the amount of parties or their dimension, meaning that it also holds for pure states representing entangled states. Let us now address a numerical approach in order to try to outperform the analytical result in Lemma 3.

\subsection{General Numerical Approach}

Given the set of two arbitrary states $P''=\{\rho_0, \rho_1\}$, we build the set of states $P=\{\rho_{\vec x} \}$ as in \autoref{eq:pbrbigger}, and we want to find out when it is possible to guarantee $\psi$-onticity ($w(\{ \mu _{0}, \mu _{1}\})=0$). In so doing, let us consider the following bound: from Lemma 1 we have that $ w(\{\mu_0, \mu_1\} )=w(\{ \mu _{\vec x} \})^{1/n}$, and from Appendix C we have the inequality $w(\{ \mu _{\vec x} \}) \leq \sigma (\rho _{\vec x})$ with $\sigma (\{\rho_{\vec x}\}):=\sum_{\vec x} {\rm Tr} \left(E_{\vec x} \rho_{\vec x} \right)$. Hence, $w(\{\mu_0, \mu_1\} ) \leq \sigma (\{\rho _{\vec x}\})^{1/n}$ such that we have an upper 
bound for the $\psi$-onticity of the original two quantum states. Since we want to achieve $w=0$, the central point is then to minimise the function 
\begin{eqnarray}
\sigma(\{\rho_{\vec x}\})^{1/n}:=\left[\sum_{\vec x} {\rm Tr} \left(E_{\vec x} \rho_{\vec x} \right)\right]^{1/n},
 \label{eq:sigma} 
\end{eqnarray}
over all the POVM measurements $m=\{ E_{\vec x} \}$ with the constraints $E_{\vec x} \succcurlyeq 0$ and $\sum_{\vec x}E_{\vec x}=\mathds{1}$. This minimisation process can be implemented, for instance, by using MATLAB with the ``YALMIP'' toolbox \cite{YALMIP}  and the ``SDPT3'' solver \cite{SDPT3}. We are then interested into the sets of states $P$ for which the numerical procedure achieves $\sigma=0$, which we will call \emph{$\sigma$-zero} regions. We point out that this numerical approach might, in principle, outperform the analytical bounds (\autoref{eq:bound}), however, as we will address later, the numerical optimisation followed here confirms the efficiency of the analytical bound in \autoref{eq:bound} \cite{PBR2012}. 

\section{Beyond the PBR Theorem}

The PBR theorem has triggered other developments. On the one hand, there have also been other results aiming to discard $\psi$-epistemic models as the  Hardy \cite{Hardy2013}, and Colbeck-Renner \cite{CR2013} theorems, however, these results also have to make use of additional assumptions \cite{ML2014}. In this regard, only the so-called \emph{maximally $\psi$-epistemic} models have been discarded without additional assumptions \cite{MPE1, MPE2, MPE3, MPE4}. On the other hand, and perhaps from a more operational point of view, the $\neg$PP compatibility property (alone) turned out to be a witness of usefulness for the task known as \emph{conclusive exclusion of states} \cite{CES2014}. Since we will be interested into the $\neg$PP compatibility of $P$ under noisy channels, we now briefly address this task, and also discuss the generalised $\rho$-onticity for the set of $n$-qubit noisy states $P$ which is going to be related with its $\neg$PP compatibility alone (without additional assumptions).

\subsection{$\neg$PP compatibility and the conclusive exclusion of states}

Let us start by describing the task of conclusive exclusion of states \cite{CES2014}. Experimentalist A (Alice) prepares a state $\phi$ out of the set of states $P=\{ \rho_i \}$ based on certain probabilities $\{ p_i \}$. She then challenges Bob to provide a state from that list (say $\rho_j$), such that  $ \phi  \neq \rho_j$. Alice further explains that Bob can perform up to one measurement on $\phi$. What can Bob do in order to increase his probability of overcoming the task? 

Let us now suppose that after knowing about $P$, $\{ p_i \}$, bob manage to prove that $P$ is $\neg$PP compatible for some measurement $m$. Let us now see how having done this, Bob can effectively overcome the task. Bob uses this $m$ measurement on $\phi$, and obtains an outcome out of the experiment, say $j$. Having obtained outcome $j$, he can now be completely sure that the state  $\phi$ could not have been $\rho_j$ ($\phi  \neq \rho_j$), and we can see that this is true by contradiction. Let us assume that $\phi = \rho_j$, since he obtained outcome $j$ we have that the probability of obtaining outcome $j$ is strictly greater than zero ${\rm Tr} ( E_j\rho_j)>0$, however, by $\neg$PP compatibility we have that ${\rm Tr} ( E_k\rho_k)=0$, $\forall k=1...2^n$ (in particular for $k=j$), which is a contradiction. Therefore, we have that if Bob manages to guarantee the $\neg$PP compatibility for $P$, he can confidently overcome the task. In fact, we can explicitly see this by considering the probability of overcoming the task which can be written as  \cite{CES2014}: $P_o=1-\sum_k {\rm Tr} (\tilde \rho_kE_k)$, $\tilde{\rho}_i=p_i\rho_i$. In our case, since $p_i=1/n$ we have that $P_o=1- \frac{1}{n}\sum_k {\rm Tr} (\rho_kE_k)=1-\frac{1}{n}\sigma$, and as we have already seen, when having $\neg$PP compatibility we get $\sigma=0$, so we indeed achieve $P_o=1$.

In addition of introducing this task, the authors in \cite{CES2014} also derived an \emph{if and only if} criterion to check the optimality of the obtained measurement $m$, from the semidefinite programming (SDP) problem looking for $\neg$PP compatibility. This criterion goes as follows. Having obtained measurement $m$, $m$ is an optimal measurement if and only if: i) $N=\sum_{i=1}^{2^n} \tilde{\rho}_i E_i$ is hermitian, where $\tilde{\rho}_i=p_i\rho_i$, and ii) $\tilde \rho_i-N \geq 0$, $\forall i=1...2^n$. With this criterion, they were able to guarantee the optimality of the measurements found by PBR in the $\neg$PP compatibility region and, additionally, the optimality of some measurements they proposed outside the $\neg$PP compatibility region. Further characterisation of this task from a communication complexity perspective has also been addressed \cite{CES2015, CES2016}.

\subsection{$\neg$PP compatibility and a slightly generalised $\psi$-ontic realism}

Another motivation for considering $\neg$PP compatibility under noisy channels is the following. In order to enquire about realism we consider a ``$\rho$-ontic'' generalisation of the $\psi$-ontic property. So far, we have addressed the $\psi$-ontic realism as a property of sets of pure states \cite{ML2014}, however, from the definition of $\psi$-ontic, it can easily be asked for $\psi$-onticity of general sets of states, this, in addition of including the previous addressed sets of pure states, it would also let us address a ``$\rho$-onticity'' for general sets of states. With this in place, we have that the $\neg$PP compatibility of the set of noisy $n$-qubit states $P$, would imply its own $\rho$-onticity.

\section{$\neg$PP compatibility of $P$ under noisy channels}

To analyse the influence of external perturbations upon the $\neg$PP compatibility of sets of quantum states, we consider the set of $n$-qubit states $P=\{ \rho_{\vec x}\}$ given by \autoref{eq:pbrbigger}, and allow them to interact with an environment such that they transform into a new set of states given by $P'=\{ \rho'_{\vec x}\}$. Since we are dealing with qubits, in order to explicitly define the transformation of the set of $n$-qubit states, we first remember the interaction of single qubits. We consider that single qubits evolve under the action of the following noise operators: bit flip $\sigma_1\equiv X$, 
phase flip $\sigma_3\equiv Z$, and bit-phase flip $\sigma_2\equiv Y$, where 
$\sigma_i$ are the Pauli matrices \cite{NC2010}. For a single qubit undergoing the action of a noisy channel, the operator-sum representation of the  
mapping on the density matrix is given by \cite{Kraus1983}: $\rho'(\theta) = \sum_{k=0}^{1}F_k\rho(\theta) F^\dagger_k =  F_0 \rho (\theta)  F^\dagger_0 +F_1 \rho(\theta)  F^\dagger_1 \label{singleF}$, where $F_0 = \sqrt{p}\,\mathds{1}, \; F_1= \sqrt{1-p} \, \sigma_i$, $p$ is the probability that the noise does \emph{not} affect the qubit, and the operators $F_k$ satisfy the completeness relation 
$\sum_{k}F_k F^\dagger_k =  \mathds{1}$. 

We next explore the $\neg$PP compatibility of the noise-affected set of $n$-qubit states. In so doing, we consider the new set $P'=\{ \rho'_{\vec x}\}$, with the new total density matrices
\begin{eqnarray}
 \rho'_{\vec x}(\theta)  = p\,\rho_{\vec x}(\theta) + (1-p)\,\sigma_{i}^{\otimes j}\otimes\mathds{1}^{\otimes(n-j)}\rho_{\vec x}(\theta)\sigma_{i}^{\otimes j}\otimes\mathds{1}^{\otimes(n-j)},
\label{eq:BSN}
\end{eqnarray}
where the noise operators are now defined as: 
$F_0:= \sqrt{p}\,\mathds{1}^{\otimes n}$ and 
$F_1:= \sqrt{1-p}\,\sigma_i^{\otimes j}\otimes\mathds{1}^{\otimes(n-j)}$, with $j=1,2,...,n$ the number of qubits affected by the noise, $p$ is the probability that the noise does not affect the qubit, and the operators $F_k$ satisfy the completeness relation 
$\sum_{k}F_k F^\dagger_k =  \mathds{1}$. Another way of considering the noise interaction, for which the $\neg$PP compatibility cannot be guaranteed (no $\sigma=0$ region), is addressed in Appendix D. We now address some analytical issues regarding the $\neg$PP compatibility of the set of states $P'=\{ \rho'_{\vec x}\}$, followed by a numerical approach toward the same goal.

\subsection{Some generalities of the noise interaction} 

There are some features regarding the new set of states given by \autoref{eq:BSN} that can be analytically addressed. Since we are looking for $\neg$PP compatibility by the minimisation of the sigma function (\autoref{eq:sigma}), let us analyse this sigma function. By placing \autoref{eq:BSN} into \autoref{eq:sigma} the sigma function reads:
\begin{eqnarray}
  \sigma(\{\rho'_{\vec x}\})^{1/n}&=&
 \Big \{p\sum_{\vec x} {\rm Tr} \left[E_{\vec x} \rho_{\vec x}(\theta) \right] +\nonumber \\
 &&  (1-p)\sum_{\vec x} {\rm Tr} \left[E_{\vec x}\sigma_{i}^{\otimes j}\otimes\mathds{1}^{\otimes(n-j)}\rho_{\vec x}(\theta)\sigma_{i}^{\otimes j}\otimes\mathds{1}^{\otimes(n-j)}\right]\Big \}^{\frac{1}{n}} ,
 \label{eq:sigmaN}
\end{eqnarray}
with $0 \leq p \leq1$ and $j=1,...,n$. Let us address some properties of this new function. 

\subsubsection{$p$-independence:}

The $\sigma$-zero region of the perturbed set of states (\autoref{eq:BSN}) is independent of the probability parameter $p$, as it is just an external factor of every trace operation involved in the optimisation of the overlap $\sigma(\{\rho'(\theta)\})^{1/n}$. As for every $p$ \autoref{eq:sigmaN} has only positive terms, all of them must go to zero separately such that the total function goes to zero. Because of this reason, without loss of generality, we have chosen $p=0.5$ for the simulations we will address later. 

\subsubsection{Lower bounds for the starting point of the $\sigma$-zero region:}

By comparing \autoref{eq:sigmaN} with \autoref{eq:sigma}, we can also expect the $\sigma$-zero region to be reduced in the noisy case (when compared to the noiseless system), due to the extra term that now requires minimisation (second term in \autoref{eq:sigmaN}), and therefore, the noiseless lower bound is also a lower bound for the $\sigma$-zero region in the noisy case.

\subsubsection{Upper bounds for the starting point of the $\sigma$-zero region:}

We now propose some upper bounds for the starting point of the $\sigma$-zero of the new set of states (\autoref{eq:BSN}) in terms of the noiseless lower bounds (\autoref{eq:bound}). Let us start by considering the three-qubit noiseless case, for which we have the measurement that minimises the sigma function (\autoref{eq:sigma}) is given by, let us say $m^3=\{E^3_{\vec x}\}$, which we now numerate them as $\{E^3_{k}\}$ with $k=1...8$. Next, we consider the four-qubit system with the noise affecting the first qubit (four-qubit one-qubit-noisy case); $F_0= \sqrt{p}\mathds{1}^{\otimes 4}$, and 
$F_1= \sqrt{1-p} \, \sigma_i \otimes \mathds{1}  \otimes \mathds{1}\otimes \mathds{1}$, 
where $\sigma_i=X, Y, Z$. The final state arising from \autoref{eq:BSN} is:
\begin{eqnarray}
\nonumber \rho'_{\vec x}(\theta) = [p\rho_{x_1}(\theta)+ (1-p)\sigma_i \rho_{x_1}(\theta) \sigma_i] \otimes \rho_{x_2}(\theta)  \otimes \rho_{x_3}(\theta)\otimes \rho_{x_4}(\theta).
\end{eqnarray}
Let us now consider a set of measurements that minimise the sigma function (\autoref{eq:sigmaN}) for this new system in terms of the previously introduced three-qubit set of measurements. By noting that from each measurement $E^3_k$ we can build two measurements as: $E^{4}_{k_1}=E^{4}_{k_2}:=\frac{1}{2} (\mathds{1}\otimes E^{3}_{k})$, we then can build the set $m^4=\{E^4_{l}\}$ with $l=1...16$. We can check that this is indeed a measurement (positive semidefinite, and complete) which is able to minimise the sigma function (\autoref{eq:sigmaN}) for the four-qubit one-qubit-noisy case. Meaning that we can use the lower bound from the three-qubit noiseless case as an upper bound for the four-qubit one-qubit-noisy case. This construction works for any $n$, and it depends on the amount of qubits affected by the noise. For instance, the four-qubit but two-qubit-noisy case will have now as upper bound the two-qubit noiseless lower bound, and similarly with other cases. If we fix the amount of affected qubits and let $n$ go to infinity, the upper bound of the whole noisy system also goes to zero, as in the noiseless case, but now at a slower rate. It is worth noting that even though these analytical upper bounds should be achieved  by the numerical optimisation, the latter might also find better upper bounds. We address these numerical concerns in what follows.

\subsection{Numerical Approach}

We start by addressing the numerical treatment for the noiseless case $(j=0)$, to then address the noisy case which we divide into individual $(j=1)$ and collective $(j>1)$ qubit-noise interaction (see \autoref{eq:BSN}): Firstly, we assume that just one qubit is affected by the action of the noise whilst the remainder qubits remain unchanged $(j=1)$. Secondly, we allow two or more qubits to interact with the noisy channels $(j>1)$. The simulations were performed up to four qubits.  Furthermore, by using the \emph{if and only if} criterion from \cite{CES2014}, we were able to check the optimality of the measurements obtained from solving the SDP problem looking for $\neg$PP compatibility. Specifically, for both single-qubit and multi-qubit noise interaction, for each measurement $M$ obtained, we checked that $M$ is an optimal measurement, or that it satisfies the criteria: i) $N=\sum_{i=1}^{2^n} \tilde{\rho}_i E_i$ is hermitian, where $\tilde{\rho}_i=p_i\rho_i$, and ii) $\tilde \rho_i-N \geq 0$, $\forall i=1...2^n$ \cite{CES2014}.

\subsubsection{Noiseless case:}

In \autoref{fig:fig3} (solid curves), we plot $\sigma (\{\rho'_{\vec x}(\theta)\})^{1/n}$ as functions of $\sin\theta$\footnote{We plot the  $\sigma^{1/n}$ functions  in terms of $\sin\theta $ instead of $\theta$  as the former is the trace distance \cite{NC2010} of the initial states $\rho_0$ and $\rho_1$ (\autoref{eq:OS}), being a 
distinguishability metric for quantum states; for $\sin \theta = 0$ we have $\rho_0=\rho_1$, and  for $\sin \theta = 1$ the states are orthogonal to each other. For two general matrices, the trace distance is defined as $\delta (\rho_1,\rho_2) :=\frac{1}{2} {\rm Tr}(|\rho_1 - \rho_2|),\, \mathrm{where}\, |A|:=\sqrt{A ^\dagger A}$. For two pure matrices, like in our case (\autoref{eq:OS}), $\delta (\rho_1,\rho_2) = \sqrt{1-\left< \psi_1 | \psi_2 \right>}=\sin \theta$.} for the noiseless case (\autoref{eq:pbrbigger} or $j=0$ in \autoref{eq:BSN}). The vertical lines in \autoref{fig:fig3} show the analytical lower bounds for the starting point of the $\sigma$-zero region (\autoref{eq:bound}) which turn out to be efficient since the results derived from the numerical approach could not outperform them (the solid curves do not achieve a wider $\sigma$-zero region). We effectively see that as $n$ increases, the $\sigma$-zero region also increases, and therefore, we can guarantee the $\neg$PP compatibility of the system. Let us now address the single-qubit noise.

\subsubsection{Single-qubit noise:}

\begin{figure}[h!]
 \center
\includegraphics[scale=0.7]{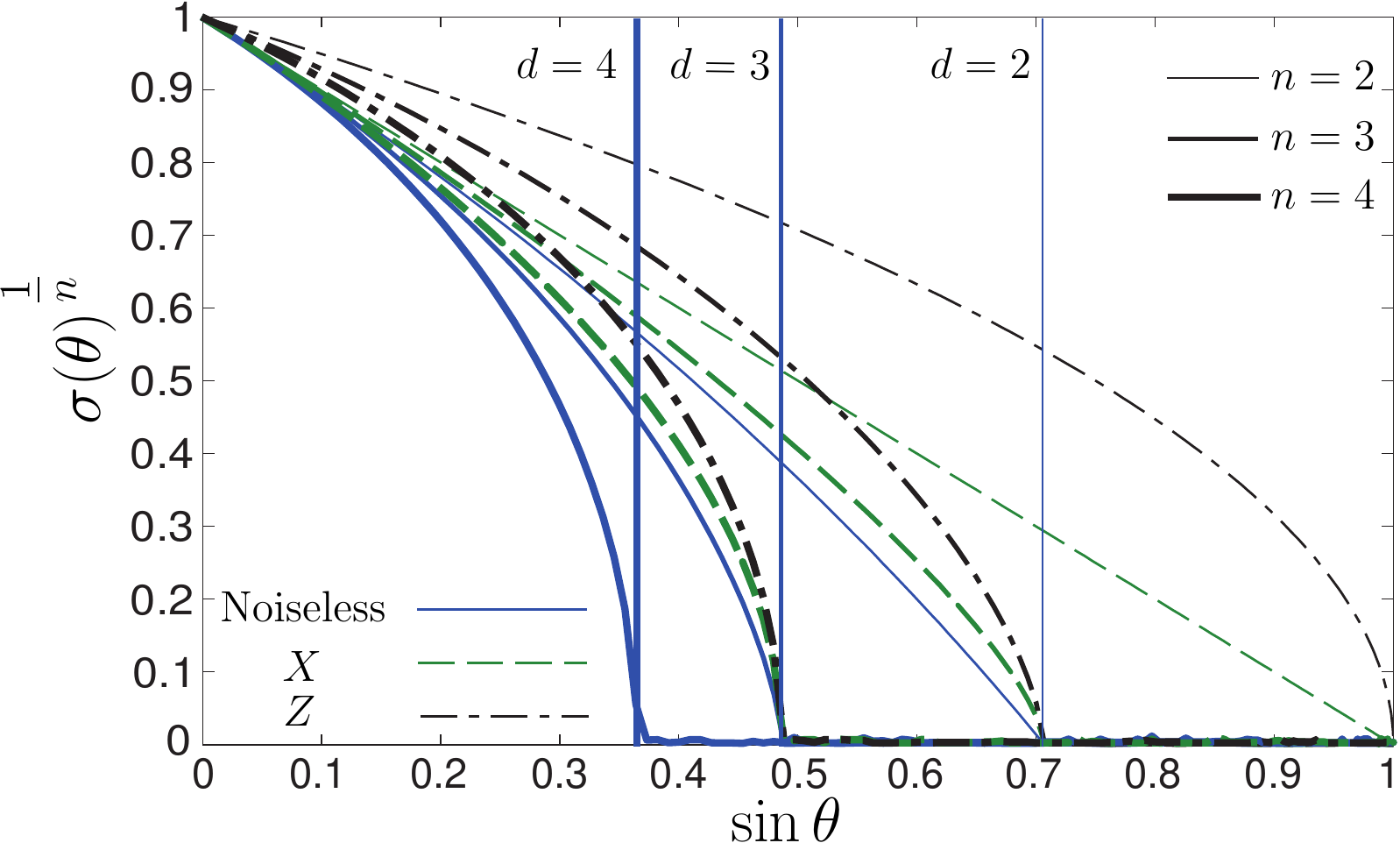}
\caption{(Color online) Sigma function $\sigma (\{\rho'_{\vec x}(\theta)\})^{1/n}$ for the set of states $P'=\{\rho'_{\vec x}(\theta)\}$ given by \autoref{eq:singleN} ($p=0.5$) as a function of  $\sin \theta$, for $X$ and $Z$ noises, and for $n=2,3,4$. Solid curves represent the noiseless cases (\autoref{eq:pbrbigger} or $j=0$ in \autoref{eq:BSN}). Vertical lines $d_n$, indicate the analytical lower bound computed from \autoref{eq:bound}: ($d_2=0.7071$, $d_3=0.4869$, $d_4=0.3740$).} 
\label{fig:fig3}
\end{figure}

The state of the $n$-qubit ensemble evolves by putting $j=1$ into \autoref{eq:BSN}:
\begin{eqnarray}
 \rho'_{\vec x}(\theta) = p\,\rho_{\vec x}(\theta) + (1-p)\,\sigma_{i}\otimes\mathds{1}^{\otimes(n-1)}\rho_{\vec x}(\theta)\sigma_{i}\otimes\mathds{1}^{\otimes(n-1)} ,
\label{eq:singleN}
\end{eqnarray}
where $F_0\equiv \mathds{1}^{\otimes n}$ and $F_1\equiv \sigma_i\otimes\mathds{1}^{\otimes(n-1)}$. We numerically minimise the sigma function (\autoref{eq:sigmaN}) for the states given by \autoref{eq:singleN}, and compare its behaviour with the one obtained in the noiseless case for two, three and four qubits. In \autoref{fig:fig3}, the dashed (dotted-dashed) curves correspond to the noisy $X$($Z$) case. In this particular single-qubit noisy scenario, we notice that the $Y$ noise produces the same effect than the $Z$ one for the three cases ($n=2,3,4$), and hence, it is not shown in \autoref{fig:fig3}. From this case we analyse the following. First, since the $\sigma$-zero region is a $p$-independent phenomenon, we present 2d plots with $p=0.5$. Second, we can see a reduced $\sigma$-zero region for the noisy cases, as expected. Third, the numerical approach for the single-qubit noisy cases could not (as in the noiseless case) find better upper bounds than the ones analytically expected. The values of $\sin\theta$ for which the overlap $\sigma (\{\rho'_{\vec x}\})^{1/n}$ completely goes to zero for the noisy cases coincide with the beginning of the $\sigma$-zero region of the $n-1$ noiseless case. For instance, the dashed and dotted-dashed curves corresponding to the $X$ and $Z$ noisy cases for $n=4$ go to zero at $\sim0.4869$ which is the value of the beginning of the zero region for the $n=3$ noiseless case. Therefore, the $\sigma$-zero region (in which it is possible to argue the $\neg$PP compatibility of the set of states $P$) is the same no matter the noise $X$, $Y$ or $Z$. We next consider the multi-qubit noise case.

\subsubsection{Multi-qubit noise:}

\begin{figure}[h!]
\centerline{\includegraphics[scale=0.8]{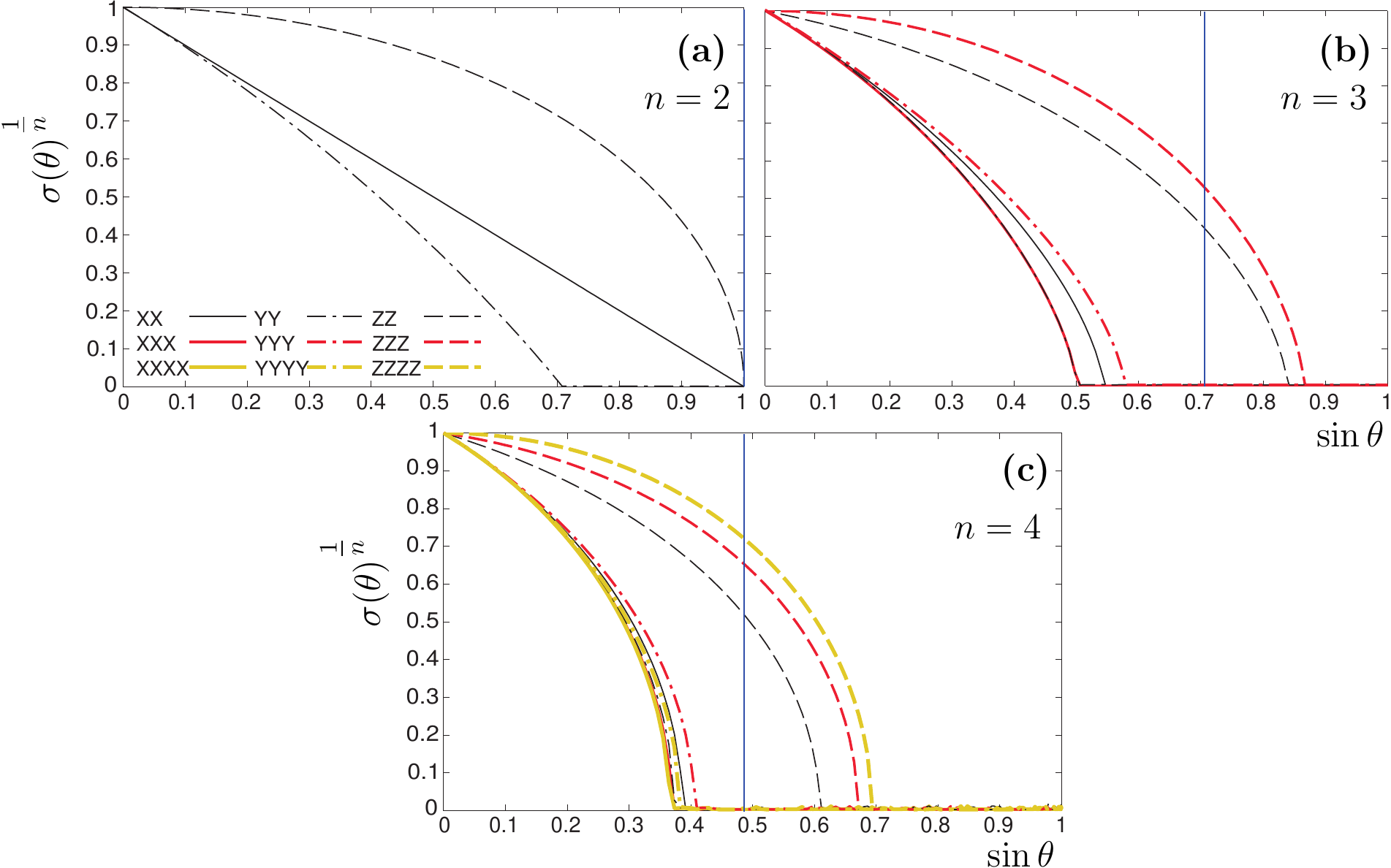}}
\caption{(Color online) Sigma function $\sigma (\{\rho'_{\vec x}(\theta)\})^{1/n}$  as a function of  $\sin \theta$, under the action of the multi-qubit noise \autoref{eq:multipleN} ($p=0.5$). (a) $n=2$, (b) $n=3$ and (c) $n=4$. The notation is as follows: for $n$ qubits, $XX$ means $X\otimes X\otimes\mathds{1}^{\otimes(n-2)}\equiv\sigma_1\otimes\sigma_1\otimes\mathds{1}^{\otimes(n-2)}$, and likewise for the other noises. The vertical lines represent the $\sigma$-zero region bound given by the corresponding single-qubit noise (see \autoref{fig:fig3}).} 
\label{fig:fig4}
\end{figure}

The state of the $n$-qubit ensemble evolves by putting $j>1$ into \autoref{eq:BSN} or explicitly:
\begin{eqnarray}
 \rho'_{\vec x}(\theta)  = p\,\rho_{\vec x}(\theta) + (1-p)\,\sigma_{i}^{\otimes j}\otimes\mathds{1}^{\otimes(n-j)}\rho_{\vec x}(\theta)\sigma_{i}^{\otimes j}\otimes\mathds{1}^{\otimes(n-j)},
\label{eq:multipleN}
\end{eqnarray}
with operators defined as: 
$F_0:= p\,\mathds{1}^{\otimes n}$ and 
$F_1:= (1-p)\,\sigma_i^{\otimes j}\otimes\mathds{1}^{\otimes(n-j)}$, and with $j=2,...,n$ the number of qubits affected by the noise. For instance, for four qubits ($n=4$) we have three possibilities for the Kraus operators, as follows:
\begin{eqnarray}
\nonumber 	F_0&=&  \sqrt{p}\mathds{1}^{\otimes 4}, \hspace{0.3cm} F_1= \sqrt{1-p} \, \sigma_i \otimes \sigma_i\otimes \mathds{1}\otimes \mathds{1},\\
\nonumber	F_0&=&  \sqrt{p}\mathds{1}^{\otimes 4}, \hspace{0.3cm} F_1= \sqrt{1-p} \, \sigma_i \otimes \sigma_i \otimes \sigma_i\otimes \mathds{1},	\\
\nonumber	F_0&=&  \sqrt{p}\mathds{1}^{\otimes 4},	\hspace{0.3cm} F_1= \sqrt{1-p} \, \sigma_i \otimes \sigma_i \otimes \sigma_i\otimes \sigma_i.
\end{eqnarray}
We report on the behaviour of these multi-qubit noise coupling as a function of the increment of the number of qubits being affected by the noise (parameter $j$ in \autoref{eq:multipleN}). We do so for two, three, and four qubits (\autoref{fig:fig4} (a), (b) and (c)). Figure \ref{fig:fig4} shows the behaviour of the sigma function (\autoref{eq:sigmaN}) under the influence of the three considered noises with different collective contributions.

\noindent
Since we have a $p$-independence phenomenon, we have taken $p=0.5$. From \autoref{fig:fig4}, we can see the obtained reduced $\sigma$-zero region, as expected. However, regarding the upper bounds, there is a complete different behaviour (in comparison with the previous noiseless and single-qubit noise cases) since the numerical simulation now does outperform the analytical upper bounds for the starting point of the $\sigma$-zero region that we had analytically proposed. Additionally, unlike the one-qubit-noisy case, these bounds are not equal, and depend on the kind of noise. We now address a couple of extra behaviours. Following the notation: for $n$ qubits, $XX$ means $X\otimes X\otimes \mathds{1}^{\otimes(n-2)} \equiv\sigma_1\otimes\sigma_1\otimes \mathds{1}^{\otimes(n-2)}$, and likewise for the other noises, the overlap $\sigma^{1/n}$ exhibits a completely opposite response when the state is affected by the 
sequence of phase flip noises (e.g., $ZZ\rightarrow ZZZ\rightarrow ZZZZ$ for four qubits) than when 
affected by the sequence of bit flip noises (e.g., $XX\rightarrow XXX\rightarrow XXXX$). In the former case ($Z$-sequence), the $\sigma$-zero region is smaller compared with that obtained for the single-qubit case, and decreases in the direction of the sequence (dashed curves in \autoref{fig:fig4}(b) and (c)). Oppositely, the bit flip $(X)$ noise sequence allows an enhancement of the $\sigma$-zero region (solid curves in \autoref{fig:fig4} (b) and (c)) that is lower bounded by the noiseless scenario (the numerical values can be seen in \autoref{fig:fig3}). Although for the bit-phase flip $(Y)$ noise sequence, the overlap function falls to zero before the corresponding single-qubit noise, it does not present a monotonic behaviour with the sequence but moves in a ``zigzag'' when the sequence goes forward (dotted-dashed curves in \autoref{fig:fig4} (b) and (c)). 

\section{Conclusion}

In this work we addressed the $\neg$PP compatibility property of the set $P$ of $n$-qubits under some noisy channels. Since this $\neg$PP compatibility is an explicit witness of usefulness for the task of conclusive exclusion of states, our results consequently address the efficiency of the task under noisy channels. In particular, based on the noiseless bounds for $\neg$PP compatibility of $P$, we proposed some analytical bounds for the noisy case which showed that for both the single-qubit and the multi-qubit noisy scenarios, it is still possible to guarantee the $\neg$PP compatibility of the final noisy $P'$, and consequently, that even the noisy $P'$ still allows for conclusive exclusion of states. Furthermore, in order to optimise these analytical results, we carried out a numerical treatment of the problem up to four qubits for which we have found the following. For the single-qubit noise interaction, the numerical optimisation could not outperform the proposed analytical bounds. In contrast, for the multi-qubit noise interaction, the numerical optimisation could now outperform the proposed analytical bounds, showing a better display of the phenomenon. Additionally, the numerical optimisations obtained after solving the SDP problem were checked to be optimal by means of the criterion from \cite{CES2014}. In other words, that the measurement $m=\{ E_i\}$ obtained from solving the respective SDP for both single-qubit and multi-qubit noise interaction satisfy that: i) $N=\sum_{i=1}^{2^n} \tilde{\rho}_i E_i$ is hermitian with $\tilde{\rho}_i=p_i\rho_i$, and that ii) $\tilde \rho_i-N \geq 0$, $\forall i=1... 2^n$. Furthermore, by adopting $\psi$-onticity as a property of general sets of states (noisy $P$ in particular), instead of being exclusively for the sets of pure states, one has that checking for $\neg$PP compatibility of a set of states $P$, leads to checking for its own $\rho$-onticity. From this point of view, the previous results would also be addressing the $\rho$-onticity of $P$ when interacting with the considered noisy channels.

\section*{Acknowledgements} 

A.F.D.  and C.E.S. gratefully acknowledge Colciencias for a ``Young Researcher'' award and a fellowship, respectively.  We thank Colciencias (grant 71003), and Universidad del Valle (grant 7930) for financial support, and the Science, Technology and Innovation Fund-General Royalties System (FCTeI-SGR) under contract BPIN 2013000100007.  

\appendix

\section{On the preparation independence postulate} 

Here we show how that PIP (\autoref{eq:PIP}) is a natural property for states like in \autoref{eq:BS} and local measurements. Given the two states $\rho_{x_i}, x_i \in \{0, 1 \}$,  and POVM $m=\{E_{k_j}\}$, by (\autoref{eq:TPL}) we have
\begin{eqnarray}
\int_\Lambda \mu_{x_i}(\lambda_{x_i})\xi(E_{k_j}|p_m, \lambda_{x_i})d\lambda_{x_i}={\rm{Tr}}(E_{k_j}\rho_{x_i}),
\label{eq:before0}
\end{eqnarray}
for the new states (\autoref{eq:BS}) we get
\begin{eqnarray}
\nonumber \int_{\Lambda^n} \mu_{\vec x}(\lambda)\xi(E_{\vec k}|p_m,\lambda)d\lambda={\rm{Tr}}(E_{\vec k}\rho_{\vec x}),
\end{eqnarray}
and assuming measurements as $E_{\vec k}=\bigotimes_{j=1}^n E_{k_j}$  (local measurements), 
\begin{eqnarray}
\int_{\Lambda^n} \mu_{\vec x}(\lambda)\xi(\vec k|p_m,\lambda)d\lambda=\prod_{j=1}^n {\rm{Tr}}(E_{k_j}\rho_{x_j}).
\label{eq:before1}
\end{eqnarray}
By replacing \autoref{eq:before0} in the right hand side of \autoref{eq:before1}, we obtain
\begin{eqnarray}
 \nonumber \mu_{\vec x}(\vec \lambda)= \mu_{x_1}(\lambda_1)...\mu_{x_n}( \lambda_{n}),
\end{eqnarray}
which is the PIP (\autoref{eq:PIP}). 

\section{Proof of lemma 1} 

Given $P=\{ \rho_{\vec x} \}$ as in \autoref{eq:BS}, and corresponding functions $\{ \mu_{\vec x}\}$, and taking into account PIP (\autoref{eq:PIP}), then the minimum over all of them reads
\begin{equation}
 \nonumber   {\rm min}_{\vec x}\{  \mu_{\vec x}(\vec \lambda) \} = {\rm min} \{\mu_{0}(\lambda_1),\mu_{1}(\lambda_1)\}\times \cdots \times {\rm min} \{\mu_{0}(\lambda_n),\mu_{1}(\lambda_n)\}. 
 \label{eq:before2}
\end{equation}
Inserting the \autoref{eq:before2} into the overlap  (\autoref{eq:overlap}), we obtain
\begin{eqnarray}
 \nonumber  w(\{ \mu _{\vec x} \})= \prod_{j=1}^n w_j(\mu _0,\mu_ {1}),
\end{eqnarray}
and due to the form of the states we are working with (\autoref{eq:BS}),
\begin{eqnarray}
\nonumber w(\{ \mu _{\vec x} \})= w(\mu_0, \mu_1 )^{n}.
\end{eqnarray}
Therefore, if the set of states $P$ (\autoref{eq:BS}) is $\psi$-ontic, then the original set of states $P''$ (\autoref{eq:OS}) is $\psi$-ontic too $\square$.

\section{Proof of lemma 2}

Let  $P=\{\rho_{\vec x} \}$ be a set of states, where the label $\vec x$ just represents a counter. 
From the total probability law (\autoref{eq:TPL}), for each $\vec x$ we have
\begin{eqnarray}
 \nonumber \int _{\Lambda^n}  {\rm min}_{\vec x}\{  \mu_{\vec x}( \lambda) \} \xi_{\vec x}(\lambda)d \lambda   \leq  \int _{\Lambda^n}   \mu_{\vec x}( \lambda)   \xi_{\vec x}(\lambda) d  \lambda,
\end{eqnarray}
with $ \xi_{\vec x}(\lambda) :=  \xi(E_{\vec x}|p_m, \lambda)$, where the second term is just the definition (\autoref{eq:TPL}). By means of \autoref{eq:BR} we obtain
\begin{eqnarray}
\nonumber \int _{\Lambda^n}  {\rm min}_{\vec x}\{  \mu_{\vec x}( \lambda) \} \xi_{\vec x}(\lambda)d \lambda   \leq {\rm Tr}(E _{\vec x} \rho_{\vec x}).
\end{eqnarray}
Adding on $\vec x$, we have
\begin{eqnarray}
\nonumber \sum_{\vec x}\int _{\Lambda^n}  {\rm min}_{\vec x}\{  \mu_{\vec x}( \lambda) \} \xi_{\vec x}( \lambda)d \lambda   \leq \sum_{\vec x} {\rm Tr}(E _{\vec x} \rho_{\vec x}).
\end{eqnarray}
We introduce the function $\sigma(\{\rho_{\vec x}\}):=\sum_{\vec x} {\rm Tr}(E _{\vec x} \rho_{\vec x})$,
\begin{eqnarray}
\nonumber \int _{\Lambda^n} \sum_{\vec x}   \left( {\rm min}_{\vec x}\{  \mu_{\vec x}( \lambda) \}\xi_{\vec x}(\lambda) \right)  d  \lambda   \leq \sigma(\{\rho_{\vec x}\}),
\end{eqnarray}
where ${\rm min}_{\vec x}\{  \mu_{\vec x}(\vec \lambda) \}$ is a function only on $\lambda$, then
\begin{eqnarray}
\nonumber \int _{\Lambda^n}  {\rm min}_{\vec x}\{  \mu_{\vec x}( \lambda) \}d \lambda  \left(\sum_{\vec x} \xi_{\vec x}( \lambda)\right) \leq \sigma(\{\rho_{\vec x}\}), \\
\nonumber \int _{\Lambda^n}  {\rm min}_{\vec x}\{  \mu_{\vec x}( \lambda) \}d  \lambda  \leq \sigma(\{\rho_{\vec x}\}),\\
\nonumber  w(\{ \mu _{\vec x} \}) \leq \sigma(\{\rho_{\vec x}\}).
\end{eqnarray}
From $\neg$PP compatibility (\autoref{eq:NPP}), ${\rm Tr}(E _{\vec x} \rho_{\vec x})  =0 \hspace{0.2cm}\forall \vec x$, then $\sigma(\{\rho_{\vec x}\})=0$, and taking into account that $w$ is a positive function, then $w(\{ \mu _{\vec x} \})=0$. Therefore the system must be represented by a $\psi$-ontic QOM $\square$.

\section{On the independent noise interaction}

In a different noise interaction scenario, we might consider the system to be perturbed by independent single-qubit noises, that is, that every qubit undergoes a map such that the total multipartite state reads:
\begin{eqnarray}
 \rho'_{\vec{x}}( \theta)=\bigotimes_{i=1}^{n}\rho'_{x_i}(\theta), \hspace{0.5cm} \rho'_{x_i}(\theta)=p\rho_{x_i}(\theta)+(1-p)\sigma_i\rho_{x_i}(\theta)\sigma_i.
 \label{eq:ind}
\end{eqnarray}
However, we found that there is no way to guarantee the $\neg$PP compatibility of the system in this scenario, as for the mixture parameter $0<p<1$, the sigma function (\autoref{eq:sigma}) never goes to zero. In \autoref{fig:fig5} we have plotted the behaviour of the sigma function for the case of $n=3$ and $\sigma_i=Z$ (similar results we obtained for the rest of ($n, \sigma_i$) configurations):

\begin{figure}[h!]
 \center
\includegraphics[scale=0.6]{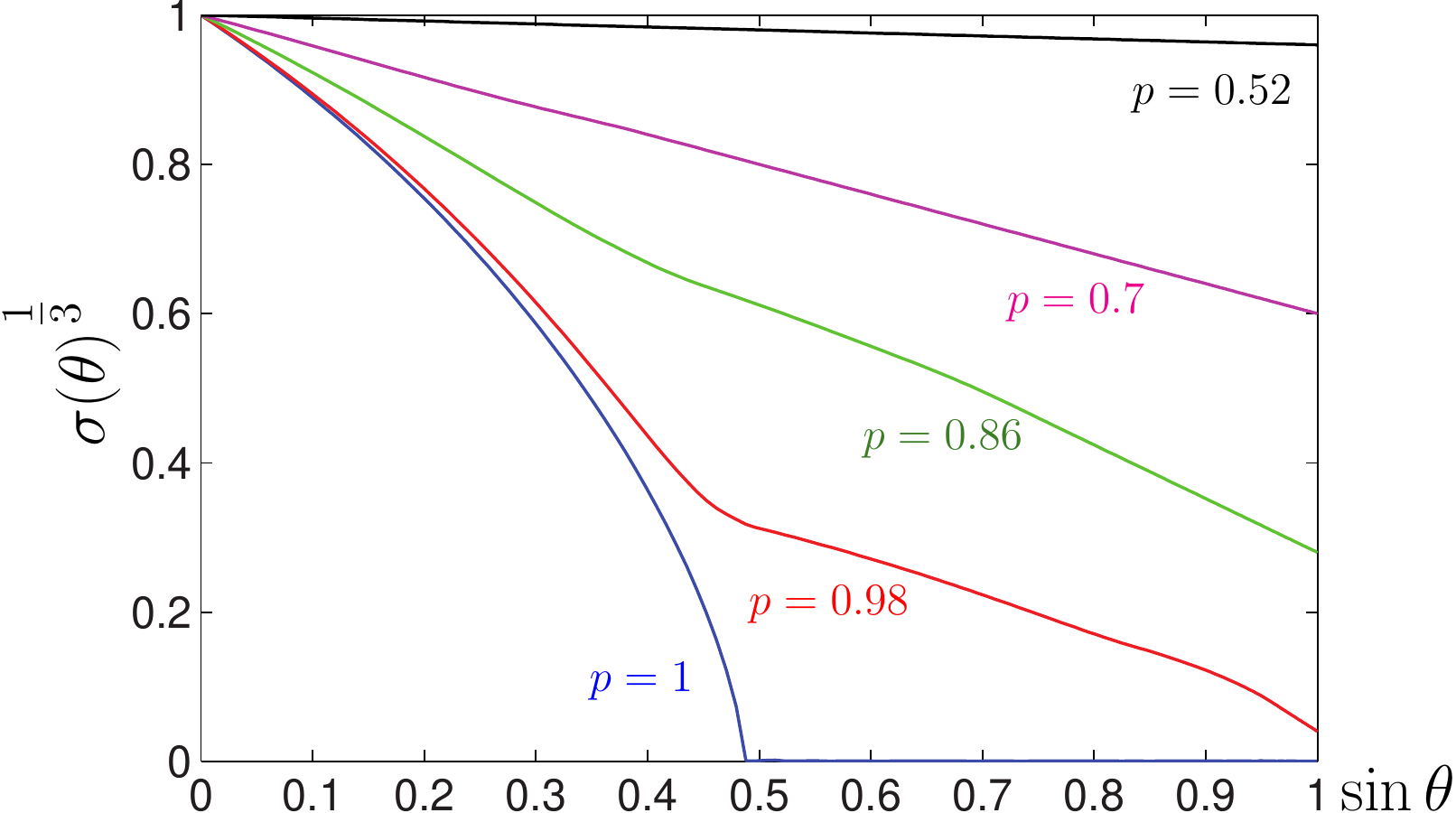}
\caption{(Color online) Sigma function $\sigma (\{\rho'_{\vec x}(\theta)\})^{1/3}$ for the set of states $P'=\{\rho'_{\vec x}(\theta)\}$ given by \autoref{eq:ind} with $\sigma_i=Z$, and for $p=1, 0.98, 0.86, 0.7$ and $0.52$, as a function of  $\sin \theta$. We can see that unlike the single-qubit, and multi-qubit noise interaction, as soon as $p<1$, we lost the $\sigma=0$ region, and therefore, the noisy set $P'$ does not allow conclusive exclusion of states.}
\label{fig:fig5}
\end{figure}

In other words, we cannot guarantee the usefulness for the task of conclusive exclusion of states of the noisy set $P'$. For the particular case where the environment completely changes the total state ($p=0$), the sigma function of the noisy state exhibits the same behaviour of the overlap computed on the original state (\autoref{eq:pbrbigger}) due to the symmetry of the superposition (\autoref{eq:OS}) with respect to the Pauli's matrices $\sigma_i$.

\newpage
\section*{References}

\bibliographystyle{unsrt}
\bibliography{bibliography.bib}

\end{document}